\documentclass[showpacs,twocolumn,showkeys,preprintnumbers,amsmath,amssymb]{revtex4}

\usepackage{graphicx}
\usepackage{dcolumn}
\usepackage{bm}
\usepackage{amsmath,amssymb,amsfonts,amssymb,amsthm}

\begin{document}


\title{Generation of optical beams with
desirable orbital angular momenta by transformation media}

\author{Weixing Shu}
\author{Dongmo Song}
\author{Zhixiang Tang}
\author{Hailu Luo}
\author{Yougang Ke}
\author{Xiaofang L\"{u}}
\author{Shuangchun Wen}\thanks{scwen@hnu.edu.cn}
\author{Dianyuan Fan}

\affiliation{Key Laboratory for Micro-/Nano-Optoelectronic Devices
of Ministry of Education, College of Information Science and
Engineering, Hunan University, Changsha 410082, China}

\begin{abstract}
We propose a scheme to controllably convert the wavefront of an
arbitrary incident beam into a helical one by compact
transformation slabs, thus enabling the output beam to carry
desirable orbital angular momentum (OAM). First, based on
transformation optics, a three-dimensional (3D) phase
transformation between any two wavefronts by flat transformation
media is established and then used to mold a wavefront of Gaussian
beam into a helical one. Second, 3D FDTD simulations are performed
to confirm the spiraling evolutions of the resultant field and
phase, clearly demonstrating OAM generated. Further theoretical
analyses show that the refractive index exhibiting a helical
distribution leads to the oppositely spiral phase front and that
it is feasible to produce desirable OAM by generators of unit OAM.
The results not only provide an additional way to manipulate phase
and photon OAM, but reciprocally shed further light on the phase
structure of helical beams, which leads to a new transformation
way by a surface.
\end{abstract}

\maketitle

\section{Introduction}\label{Introduction}

Phase is an important characteristic of electromagnetic waves. As
well known a beam with a phase $\exp{(il\theta)}$ depending on the
azimuthal angle $\theta$ and an integer $l$, typically
Laguerre-Gaussian (LG) mode, possesses helical wavefront or
optical vortex.  It thus has an azimuthal component to the linear
momentum, resulting in an orbital angular momentum (OAM) of
$l\hbar$ per photon along the beam axis
\cite{Allen1992,Soskin1997,Curtis2003}. Owing to its fascinating
properties, the beam has received a great deal of attention
\cite{Allen2003,Soskin2001,Yao2011} and has brought novel
applications in manipulating particles \cite{Grier2003} or atoms
\cite{Andersen2006}, classical \cite{Gibson2004} or quantum
communication \cite{Molina-Terriza2007}, imaging
\cite{Torner2005}, optical data storage \cite{Voogd2004} as well
as biophysics \cite{Dholakia2011}. Among the usual methods to
produce helical beams: mode converters require specific input
modes \cite{Allen1992}; spiral phase plates have special profiles
difficult to fabricate especially in optical frequencies and to
obtain high quality beam \cite{Beijersbergen1994,Sueda2004};
computer-generated holograms similar to diffraction gratings have
limited conversion efficiency and mode purity \cite{Bazhenov1992};
spatial light modulators made of liquid crystal pixels are a
tunable generation way, yet need be electronically driven
\cite{Curtis2003}. The last two require low input power as well.

To harness  phase, transformation optics that allows the control
of waves in a desirable way provides a promising route
\cite{Pendry2006}. It translates electromagnetic behaviors wanted
into the distribution of material parameters that are then
implemented by transformation media, usually metamaterials
\cite{Cui}. This controllable method provides great flexibility in
designing optical devices and a variety of exciting functions have
been proposed \cite{Schurig2006,Chen2010}. To manipulate phase
recent work realized field rotation \cite{Chen2009} and directive
emission that has important applications such as highly directive
antennas \cite{Jiang2008,Kwon2008a,Lin2008a,Tichit2011,Kong2007}.
The latter was achieved by transforming curved waves into plane
ones directly. That transformation way leads to devices with
irregular profiles to perform the conversion between curved waves
\cite{Ma2008a} or even between a plane wave and another one
deflected \cite{Yu2011}.  However, flat configurations are
preferred in practical applications,  especially as plug-and-play
devices, being compact enough and free of aberrations
\cite{Kundtz2010}. To obtain flattened devices further
transformations need be carried on geometrical shapes
\cite{Roberts2009,Yang2011}. Instead we proposed to directly
accomplish phase conversion between waves by flat transformation
media \cite{Shu2012}.

According to Fermat's principle the distribution of refractive
index determines the generated wave phase \cite{Hecht2002}. So
far, metamaterials of refractive index ranging from positive to
negative or from high to zero have been fabricated successfully
\cite{Khoo2006,Pimenov2006}, while transformation optics permit to
manipulate phase exactly. By combining them it is hence possible
to convert any input modes into pure helical ones of any orbital
angular momenta (OAMs). Moreover, the large range of material
parameters available will make the design ultra-compact whereby
generating desirable OAMs as conveniently and economically as
plug-and-play devices.

In this work we hence propose a controllable scheme based on
transformation optics to produce helical phase fronts by compact
flat configurations. We first put forward a method of
three-dimensional (3D) transformation between any two wavefronts
and use it to shape the helical wavefront out of Gaussian beam.
Next we perform 3D FDTD simulations to reveal the physical process
and confirm the generation of OAM. The phase structure of light
beams with OAM are examined from the viewpoint of transformation
optics, which leads to a new transformation way by a surface.
Further we propose available materials to realize the design and
demonstrate to produce arbitrary OAMs tunably using generators of
unit OAM. The device is theoretically reflectionless because the
3D transformation achieves the impedance match
\cite{Emiroglu2010}, so the conversion efficiency is high. If
proper materials employed the slab may support high intensity
beam.


\section{Principle}\label{formulation}

\subsection{Light beam with helical wavefronts}

\begin{figure}
\centering
\includegraphics[width=5cm]{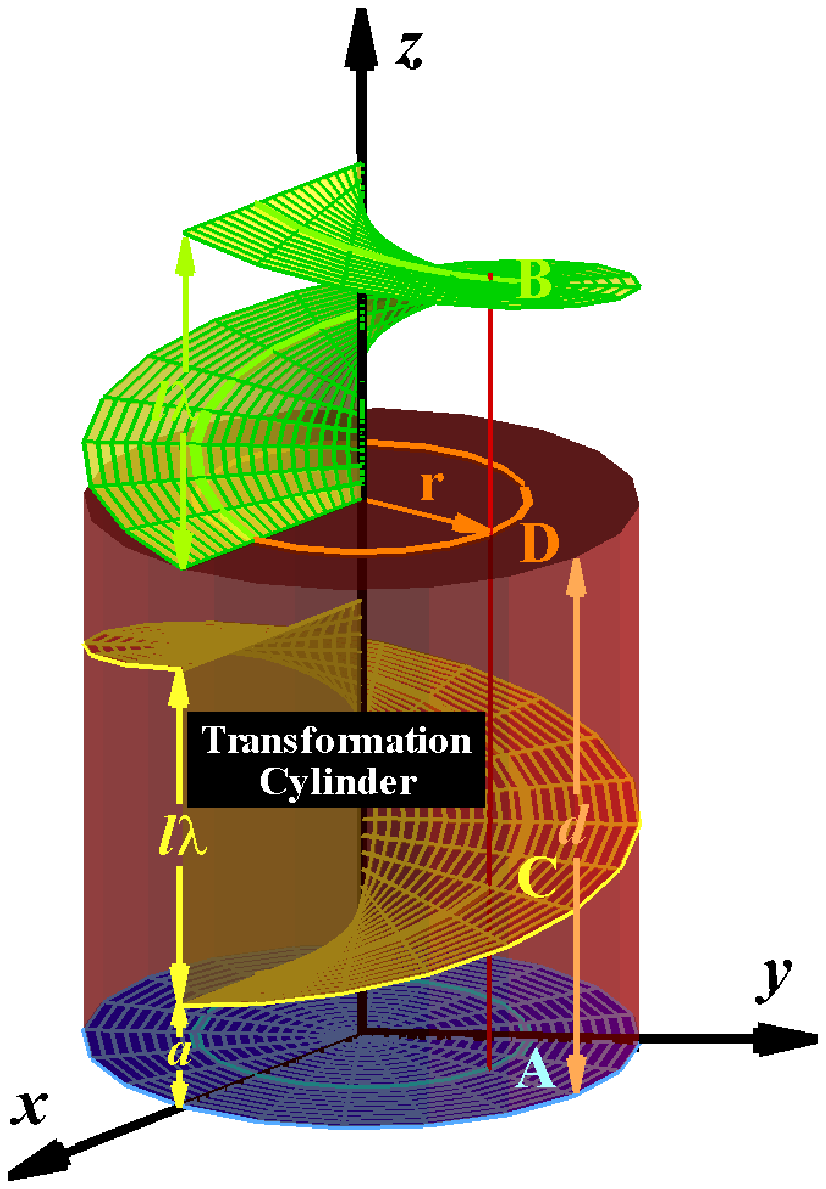}
\caption{(color online). Schematic of the coordinate
transformation upon which the resultant transformation cylinder
converts a wavefront of Gaussian beam into a helical one. The
common way is to transform the former (bottom blue surface) into
the latter (top green helicoid) directly, i.e. A$\rightarrow$B,
whereby the resultant device has an irregular profile.
Alternatively we employ a new way that is to transform a surface
(lower yellow helicoid) opposite to the wavefront desired (green
helicoid) into a plane (upper red surface), i.e. C$\rightarrow$D,
which leads to a flat cylinder.}\label{space}
\end{figure}

For an initially Gaussian-enveloped beam bearing
OAM, the field is an eigenmode of the paraxial Helmholtz equation
\cite{Allen1992,Soskin1997,Curtis2003}:
\begin{eqnarray}\label{field}
E_{l}(r,\theta,z)=E_0\frac{w_0}{w}\left(\frac{r}{w} \right)^{|l|}
\exp{\left(\frac{-r^{2}}{w^2}\right)}\exp{[-i\Phi(r,\theta,z)]}
\end{eqnarray}
where $E_0$ is the amplitude parameter, $w_0$ is the waist radius,
the beam width $w=w_0\sqrt{1+(z/z_R)^2}$, the Rayleigh length
$z_R={\pi}w_0^2/\lambda$, the wavenumber $k=2\pi/\lambda$,
$\lambda$ denotes the vacuum wavelength, the phase
\begin{eqnarray}\label{phase}
\Phi(r,\theta,z)=-(|l|+1)\tan^{-1}\frac{z}{z_R}+\frac{kr^{2}}{2R(z)}
+l\theta+kz,
\end{eqnarray} and $R(z)=z+z_R^2/z$ is the wavefront
radius of curvature. This is actually $\hbox{LG}_0^l$ mode that
reduces to a fundamental Gaussian beam when $l=0$
\cite{Allen1992}. In Eq.~(\ref{phase}) the first term and $R(z)$
are relatively slowly varying functions and are effectively
constants within the beam width on each wavefront. Then the
equiphase surface, $z=g_H(r,\theta)=c_H-{r^{2}}/{2R}-l\theta/k$,
defines a helix circling the $z$-axis with a pitch of $l\lambda$,
while with a radius of curvature $R$ for a fixed $\theta$.
Meanwhile the wavefront of Gaussian beam
$z=g_G(r,\theta)=c_G-{r^{2}}/{2R}$ is a paraboloidal surface and
$c_H$ and $c_G$ are constants.  As shown in Fig.~\ref{space} the
above green helicoid corresponds to helical beam with $l$, while
the low blue surface Gaussian beam.

\subsection{3D phase transformation method by slabs}

To convert an arbitrary wavefront into another, e.g. $z=g_G$ into
$z=g_H$, one can transform points on the former into the latter
immediately (e.g., $A\rightarrow  B$). The phase difference added
is $k(g_H-g_G)$. Such an intuitive and direct method, however,
leads to the device with an irregular shape \cite{Ma2008a}.
Alternatively we introduce a new way in this work and consider a
flat cylindrical design of length $d$ and radius $r_0$. We find
that if a profile of the distance between the two wavefronts is
transformed into a plane surface $z=d$ (e.g., $C\rightarrow  D$),
the phase difference introduced equals $k[d-(g_G-g_H)]$, the same
as the former except a constant which will not affect the phase
profile. According to the principle of equal optical path length
\cite{Hecht2002}, the two transformation ways will generate the
same wavefront, whereas the latter results in a \textit{flat}
transformation medium \cite{Shu2012}. The 3D transformation may be
written as
\begin{eqnarray}\label{transform cylindrical}
r'=r,~\theta'=\theta,~z'=dz/\Delta,
\end{eqnarray}
where $\Delta=g_G-g_H$ refers to the spatial separation.
Following transformation optics \cite{Pendry2006} the permittivity
and permeability tensors of the transformed medium
{\boldmath$\varepsilon$} and {\boldmath$\mu$} are respectively
related to the original {\boldmath$\varepsilon_o$} and
{\boldmath$\mu_o$} by
$\boldsymbol{\varepsilon}=\Lambda\boldsymbol{\varepsilon_o}
\Lambda^T/\hbox{det}(\Lambda)$ and
$\boldsymbol{\mu}=\Lambda\boldsymbol{\mu_o}\Lambda^T
/\hbox{det}(\Lambda)$ where $\Lambda$ is the Jacobian matrix
between the transformed and original coordinates. Upon the above
transformation we obtain for the cylinder
\begin{eqnarray}\label{eu cylindrical}
\boldsymbol{\varepsilon}(\boldsymbol{\mu})=\left [
\begin{array}{ccc}
\frac{\Delta}{d} & 0 & 0\\
0 & \frac{\Delta}{d} & -\frac{z\Delta'}{d r} \\
0 & -\frac{z\Delta'}{d r} &
\frac{d}{\Delta}+\frac{{z}^2{\Delta'}^2}{d{r}^2\Delta}
\end{array} \right ],
\end{eqnarray}
where $\Delta'=\partial{\Delta}/\partial{\theta}$ and the original
space is vacuum.

\subsection{Application to produce helical beam}

Applying the above method we convert an incident wavefront $g_G$
located at $z=a$ into the exit one $g_H$ twisting in the
$-\hat{\theta}$ direction with a phase variation $2\pi l$ in one
around.  Then $\Delta=a+l\theta/k$, the surface spiraling in the
$\hat{\theta}$ direction (middle yellow helicoid), is to be
transformed into a plane surface $z=d$ (top red surface). Here the
virtual space is the domain below the helicoid, whereas the
physical space is the cylinder \cite{Li2008,Tang2010}. For
convenience we rewrite the above results in the Cartesian
coordinate system $Oxyz$. The transformation now becomes
\begin{eqnarray}\label{transform cartesian}
x'=x,~y'=y,~z'=2\pi z/(n\theta+2\pi m),
\end{eqnarray}
where $\theta=\tan^{-1} (y/x)$ is the azimuth angle in the $Oxy$
plane, $n=l\lambda/d$ and $m=a/d$. Note that $a$ related to the
initial point coordinate is introduced to avoid singular points in
$\boldsymbol{\varepsilon}(\boldsymbol{\mu})$. Because the
transformation is carried out in the $z$ direction, it is feasible
to consider the longitudinal coordinate independent of transverse
ones, leading to anisotropy reduced greatly
\cite{Li2008,Tang2010}. Then Eq.~(\ref{eu cylindrical}) becomes
diagonal,
\begin{eqnarray}\label{eu cartesian}
\boldsymbol{\varepsilon}(\boldsymbol{\mu})=\hbox{diag}
\left[m+\frac{n\theta}{2 \pi}, m+\frac{n\theta}{2 \pi},\frac{2
\pi}{n\theta+2m\pi}\right].
\end{eqnarray}
As a result, the impedance match is achieved on the boundary, so
the cylinder can be regarded reflectionless \cite{Emiroglu2010}
and the conversion efficiency will be high.

\section{Numerical confirmation}\label{generation}

\subsection{Transverse field and phase}

\begin{figure}
\centering
\includegraphics[width=4cm]{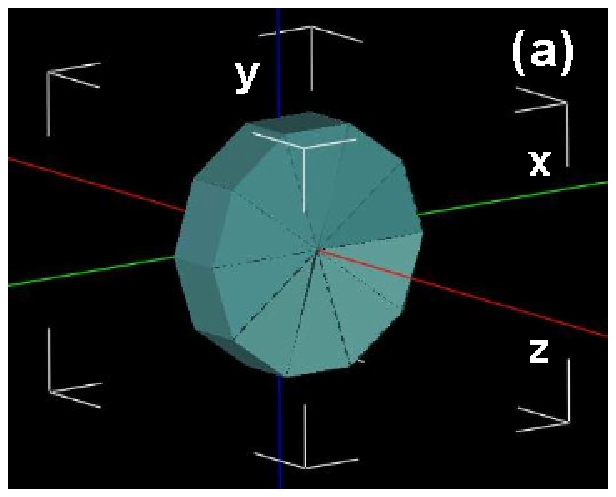}
\includegraphics[width=4cm]{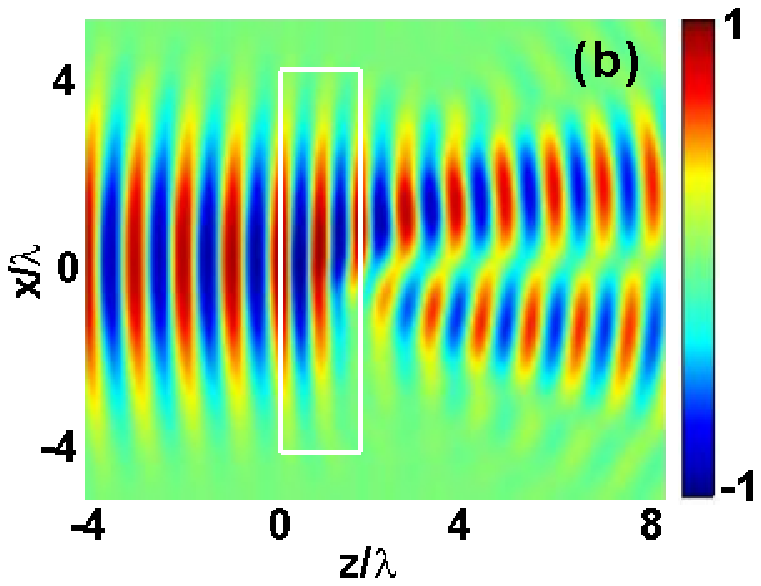}
\caption{\label{cefield}(color online). (a) Discrete model to
simulate. (b) Normalized field magnitude across the longitudinal
section of the cylinder.}
\end{figure}

We now perform numerical simulations to validate the theoretical
results and to reflect the physical process involved using the 3D
finite-difference time-domain (FDTD) methods \cite{EastFDTD}.
Consider a $y$-polarized Gaussian beam with $w_0=2\lambda$
normally incident on a cylinder of $d=2\lambda$, $r_0=4\lambda$,
$a=2\lambda$, and $l=1$. We take $\lambda=1500$nm for the sake of
simplicity and feasible implementation
\cite{Valentine2009,Gabrielli2009}. Considering practical
realization, the cylinder is divided into $10$ sectors of discrete
$\boldsymbol{\varepsilon}(\boldsymbol{\mu})$ as illustrated in
Fig.~\ref{cefield}(a). The field distribution in the longitudinal
section across $\theta=0$ and $\pi$ is shown in
Fig.~\ref{cefield}(b). Apparently the outgoing field vanishes
along the propagation axis indicating a phase singularity, a
characteristic inherent to helical beams \cite{Allen2003}. One can
see that $a/\lambda$ equals the period number of wave in the lower
section $\theta=0$. Although it may affect the constitutive
parameters and the field within the cylinder, $a$ would not change
the output field or OAM.

\begin{figure}
\centering
\includegraphics[width=4cm]{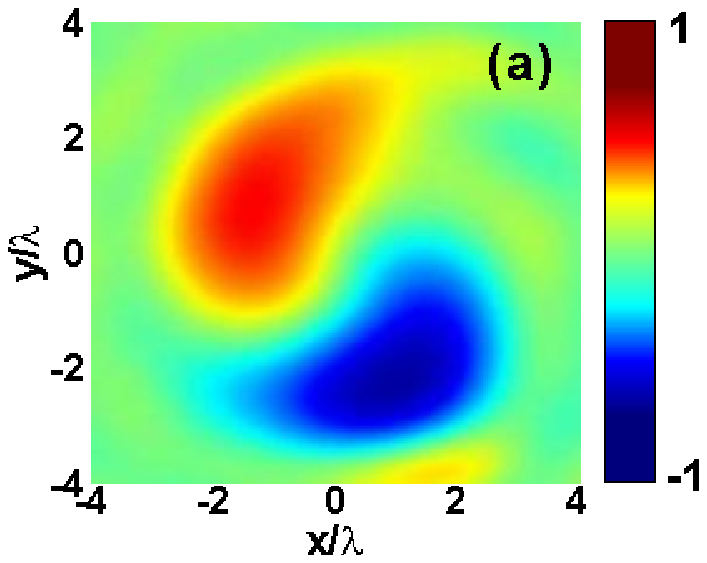}
\includegraphics[width=4cm]{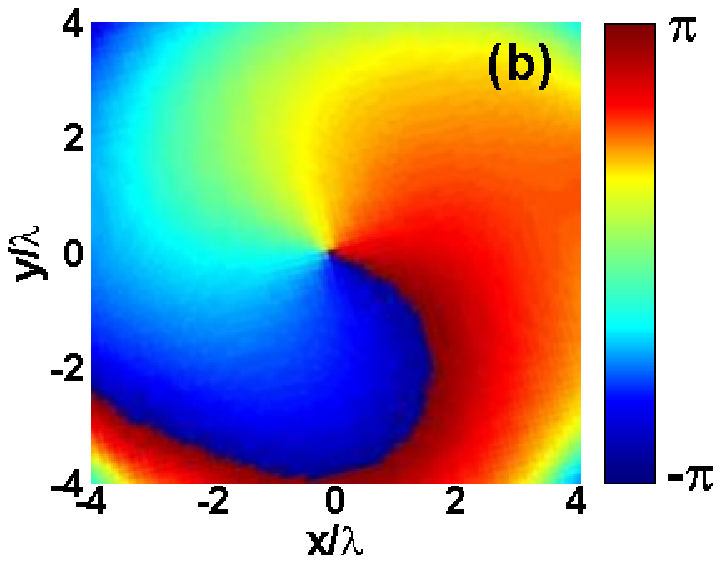}
\includegraphics[width=4cm]{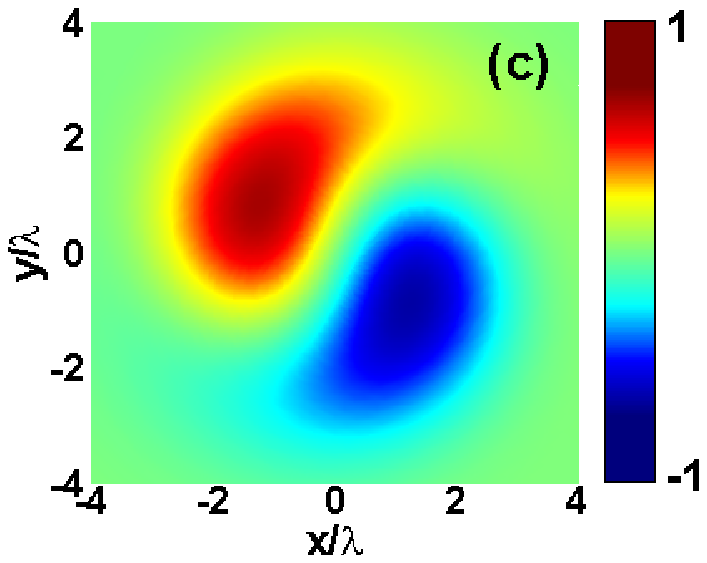}
\includegraphics[width=4cm]{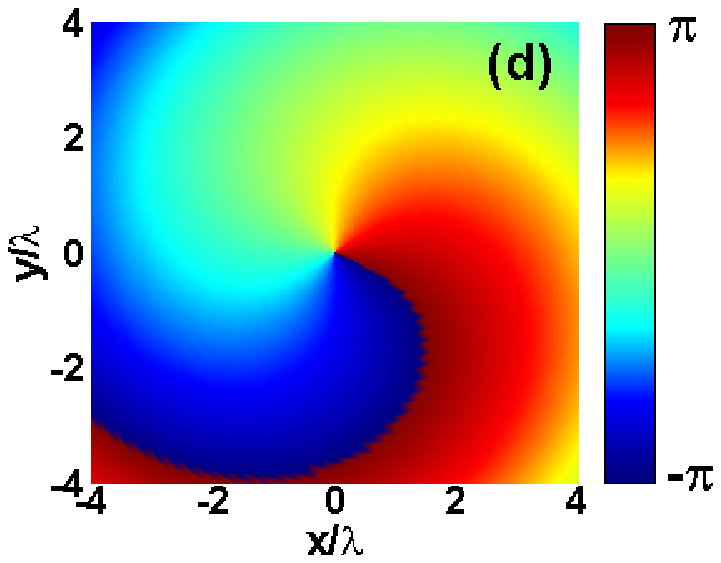}
\caption{\label{hengfieldphase}(color online). Transverse (a)
field and (b) phase distributions of the transmitted beam
$5\lambda$ away from the exit surface. (c) and (d) are the
theoretical results determined by Eq.~(\ref{field}) corresponding
to (a) and (b), respectively.}
\end{figure}

Further we show the instantaneous field and phase of the output
beam in a cross section in Fig.~\ref{hengfieldphase}. Apparently
the field distribution displays two spiral arms moving
anticlockwise. At the same time, a phase change of $2\pi$ occurs
in a clockwise round. That indicates the resultant wavefront
consists of $|l|=1$ helicoid. The simulated results in (a) and (b)
agree satisfactorily with the theoretical ones by
Eq.~(\ref{field}) in (c) and (d), respectively. So the photon has
acquired an $\hbar$ OAM. The OAM does not be generated from
nihility, but involves exchange of momenta between the beam and
the cylinder, while the total OAMs are conserved. Usually this
leads to measurable mechanical consequences
\cite{Allen1992,Allen2003}.

\subsection{Distribution of refractive index}

\begin{figure}[htbp]
\centering
\includegraphics[width=4cm]{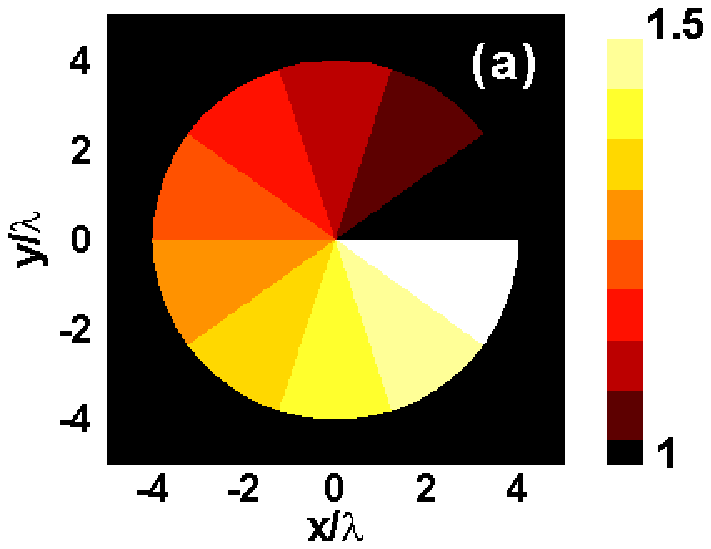}
\includegraphics[width=4cm]{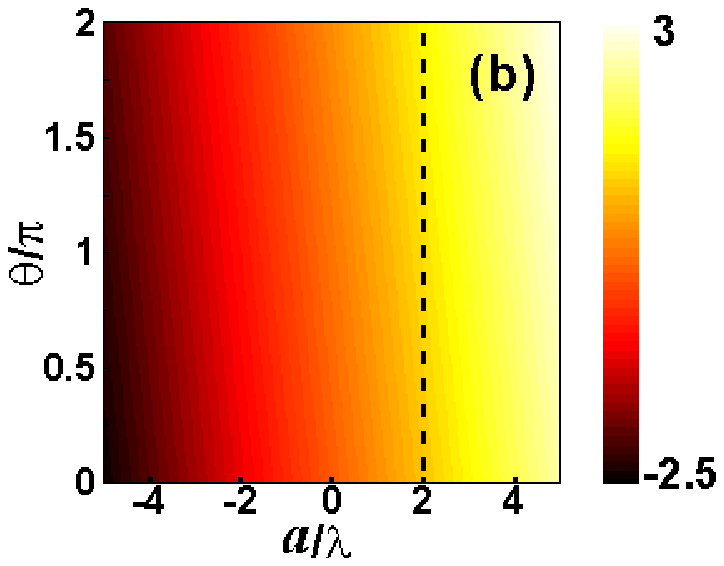}
\caption{\label{helical_n}(color online). Refraction index
distribution (a) in the transverse section and (b) around the
circumference for different $a$.}
\end{figure}

From the above the complex amplitude transmittance of
electromagnetic wave through the slab can be written as
$t=\exp{(-ik\Delta)}$ \cite{Goodman1996}. Then we introduce the
effective average refractive index to characterize the slab as
usual \cite{Li2008,Tang2010}, $n_{eff}=\Delta/{d}$. It is
consistent with $n_{eff}=\sqrt{\varepsilon_y\mu_x}$ deduced from
Eq.~(\ref{eu cylindrical}) for $y$-polarization. Using
Eq.~(\ref{eu cartesian})
\begin{equation}\label{n}
n_{eff}=m+n\frac{\theta}{2\pi},
\end{equation}
so the refraction index is helically distributed, as shown in
Fig.~\ref{helical_n}(a) for the slab in Fig.~\ref{cefield}.
Obviously the anticlockwise helical $n_{eff}$ gives rise to a
clockwise helical phase front in Fig.~\ref{hengfieldphase}(b) and
then an anticlockwise helicity on the ray trajectory in
Fig.~\ref{hengfieldphase}(a), as anticipated.  The azimuthal
distribution of $n_{eff}$ for different choices of $a$ is shown in
(b) where the dashed line corresponds to the slab in
Fig.~\ref{cefield}. It shows that $n_{eff}$ increases with $a$,
i.e. the extent of coordinate compression. Accordingly the choice
of transformations determines whether $n_{eff}$ or
$\boldsymbol{\varepsilon}$($\boldsymbol{\mu}$) are high or low,
positive or negative, or near zero.

\subsection{Physical process of OAM generation
            and the phase structure of helical beam}

The underlying physics of OAM generation can be studied even
further by viewing the movies S1 to S3, corresponding to
Fig.~\ref{cefield}, Fig.~\ref{hengfieldphase}(a) and (b),
respectively \cite{Movies}. S1: As the wave propagate through the
cylinder, the wavefront begins to advance in the section of larger
$\theta$ with larger $n_{eff}$, exhibiting a compression in
wavelength. Then a phase difference of $l\pi$ is produced between
the upper and lower parts of a longitudinal section, yet the phase
on the central $z$ axis becomes undetermined. S2 and S3: Through
the cylinder $t=\exp{(-ika-il\theta)}$ is imprinted onto the field
of the output beam, resulting a helical wavefront. Evolving with
time, the output transverse field (S2) and phase (S3) will rotate
round $z$ axis. \emph{The transverse phase for any $z$ changes
continuously by $2\pi l$ around the beam circumference, while the
center is a phase singularity} (we denote this fact by $\ast$).
These results agree with the theory about OAM beams, confirming
the above principle based on transformation optics.

Reciprocally the principle from the viewpoint of transformation
optics reveals the dynamic phase structure of beams with OAM.
Since the wavefront is defined as a equal-phase surface,
\textit{there is one wavefront traversing any point of the
transverse phase profile} in Fig.~\ref{hengfieldphase}(b) or S3.
Following ($\ast$) each wavefront will rotate, too and undergoe a
continuous transition into the next sheet after one round trip
\cite{Soskin1997}. Also by the above principle we conclude that
\emph{one wavefront is a continuous helicoid with a pitch
$l\lambda$} (we denote the two statements in italics by
$\ast\ast$).

It has been pointed out in literature that the wavefront structure
is composed from $|l|$ identical helices nested on the beam axis
separated by $\lambda$ \cite{Allen2003}. This perception
(designated by $\ddagger$) was adopted in order to interpret the
fact (indexed by $\star$) that when such a helical beam interferes
with a plane wave, it produces a transverse intensity pattern with
$|l|$ spiral arms \cite{Padgett2004}. Now ($\ast$) and
($\ast\ast$) provide an alternative perspective to understand
($\star$). According to ($\ast$) as shown in
Fig.~\ref{hengfieldphase}(b) or S3, there always exist $|l|$
points in-phase with the plane wave wherein arises the total
constructive interference \cite{Hecht2002}. Also due to the
azimuthal dependence and the radial curvature of phase, the
resultant intensity assumes $|l|$ spiral fringes. Undoubtedly,
corresponding to the $|l|$ points there exist $|l|$ wavefronts
which are just those in ($\ddagger$), yet far from the whole.
Based upon ($\ast$) and ($\ast\ast$), \emph{the overall wavefronts
consist of infinite sequential helicoids with a uniform pitch
$l\lambda$ yet individual relative phases in $[0,2\pi l]$
separated by infinitesimal distance on the beam axis}.

So it is reasonable to adopt a physical picture combining ($\ast$)
and ($\ast\ast$) to describe the dynamic structure exactly and
concisely, by which one can find two coordinate transformation
ways to realize helical beams. This first based on ($\ast\ast$) is
to carry 3D space transformation (3DST), which is just the content
of the present work. The other based on ($\ast$) means a helical
beam can be obtained through a two-dimensional (2D) \emph{surface}
transformation (2DST). It indicates that the design also can be
implemented by a \emph{surface}. Most recently Yu \emph{{et al}}
produced vortex beam by a 2D array of subwavelength resonators
\cite{Yu2011S}. The 2DST method coincides with their result which
followed a different theory though, indicating the 2DST feasible.
As being unique the 2DST may become a significant approach to
phase transformations, especially to obtain ultra-compact devices.
Recently another type of 2D transformation was used to control
surface electromagnetic waves \cite{Huidobro2010,Liu2010}, which
was essentially an application of the 3D transformation to 2D wave
problems. By contrast, the 2DST is to control 3D waves by 2D
transformation and materials.

\subsection{Generation of arbitrary OAMs and feasible
physical realizations}

\begin{figure}
\center
\includegraphics[width=5cm]{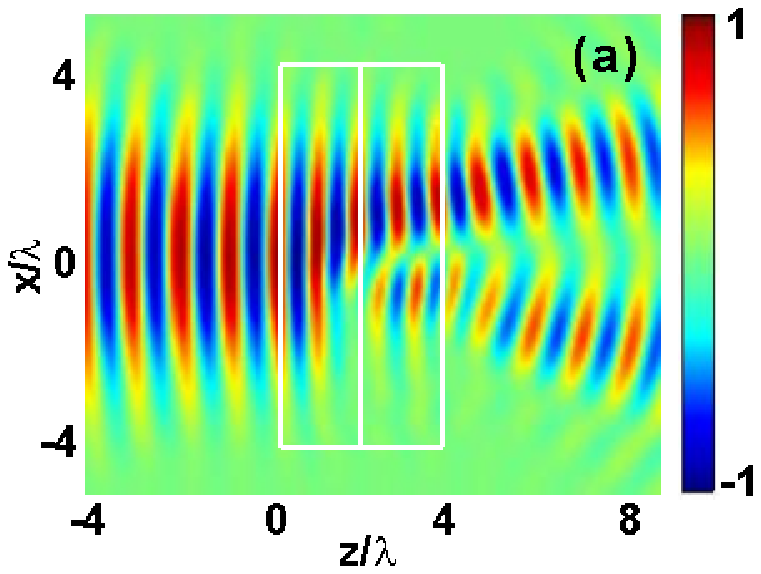}
\includegraphics[width=3cm]{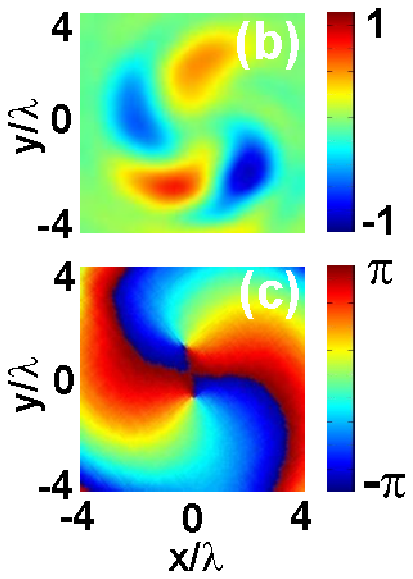}
\caption{\label{cefieldqu}(color online). Distributions of (a) the
field in the longitudinal section of two cylinders, the exit (b)
field and (c) phase in the transverse plane $5\lambda$ away from
the exit surface.}
\end{figure}

It follows from Eq.~(\ref{n}) that one can realize the same
helical beam as Fig.~\ref{hengfieldphase} but by isotropic
all-dielectric media. To show this result and to demonstrate the
flexibility of the compact design, we use two such slabs to
generate helical beam with $l=2$. Similarly $n_{eff}$ takes $10$
discrete values.  The simulated results are shown in
Fig.~\ref{cefieldqu}. In (a) the phase in the lower half section
is $2\pi$ advance against that in the upper part outside the
cylinder, coincident with (b). Extra care is required in
understanding (c). It seems like the phase consisted of two
identical segments from $-\pi$ to $\pi$. In fact the phase should
evolve continuously with a change $4\pi$ in one round according to
(*). The specious appearance in (c) is due to the restriction of
the mathematical software on the argument of the complex number
which is always between $-\pi$ to $\pi$. (b) and (c) agree well
with the theoretical results by Eq.~(\ref{field}) for a helical
beam with $l=2$. Thus two pieces of unit OAM generator in (a) have
imparted an OAM of $2\hbar$ on the output beam. Meanwhile, we find
from simulations that when $n_{eff}$ is discretized more, even
into $3$ values, helical wavefronts still appear. So the
discretization process will not affect the generation of OAM
fundamentally, which greatly facilitates the realization of the
design. Then one can obtain any OAMs in such a superposing way.

The above refractive index distribution can be realized by a
dielectric slab with air columns or vice versa whose size or
density determine $n_{eff}$ by the effective medium theory
\cite{Valentine2009,Gabrielli2009}. What's more, if the dielectric
medium is chosen with a high threshold of damage, e.g. silica, the
slab may be able to support high-power beam \cite{Sueda2004}.
Recent developments also make such an OAM generator practically
realizable. Metamaterials with material parameters radial,
azimuthal and both dependent have been constructed in
electromagnetic cloaks \cite{Schurig2006}, field rotation
\cite{Chen2009} and carpet cloak
\cite{Valentine2009,Gabrielli2009}, respectively. Hence, there
essentially does not exist technological obstacle to actualize the
parameters of Eq.~(\ref{eu cylindrical}).  As the loss is
inevitable in metamaterials, it affects the amplitude and phase at
individual points, but the overall relative phase can be kept
helical. So the loss will not prevent the generation of OAM though
may affect the quality of beam, as confirmed by additional
simulations. Particularly it will be out of question when
all-dielectric media are used.

It should be noted that recently a new approach has been proposed
to generate OAM beams using inhomogeneous anisotropic media with
the configuration very similar to that in Fig.~\ref{helical_n}
\cite{Machavariani2008}. The media could be implemented by
anisotropic crystals \cite{Machavariani2008,Volyar2006}, liquid
crystals \cite{Marrucci2006}, or subwavelength gratings
\cite{Biener2002,Niv2005}. Therein the phase was changed through
manipulating the polarization. The added phase results from the
so-called Pancharatnam-Berry geometric phase that accompanies
space-variant polarization manipulation
\cite{Biener2002,Niv2005,Milione2012}. From the viewpoint of
momentum conversion, it is the spin angular momentum of the input
beam that is transferred into the OAM of the output beam. In
contrast, the method reported here is to transform the phase
immediately, without involving the polarization. The phase is
introduced through optical phase differences. So the exchange of
OAM takes place between the output beam and the transformation
medium.

\section{Conclusion}

In summary, we have shown it is possible to enable a beam to carry
OAM employing transformation media. Applying a 3D phase
transformation method, we obtain the material parameters with a
helical distribution of refraction index. Further 3D FDTD
simulations reveal the dynamic process of OAM generation, confirm
the theoretical result and manifest the feasibility to obtain any
OAMs. The results not only present a new route to accurately
produce OAM, but reciprocally reveal the properties of beams
bearing OAM. We recognize that it is reasonable to adopt a
physical picture combining ($\ast$) and ($\ast\ast$) to describe
the dynamic structure of phase exactly and concisely. Such an
understanding allows a helical beam to be obtained by two
transformation ways, the 3DST discussed presently and the 2DST
which results in a planar device, i.e. implemented by a surface.
It is hoped to return to the latter issue elsewhere.

Besides generating OAM beams the cylinder permits to detect a
particular OAM state \cite{Allen2003,Weihs2001}. When operated in
reverse, the cylinder for $l$ flattens the helical phase fronts of
$-l$-beam  and the output beam, now with planar phase fronts, can
be detected. This allows the $l$ value of any helical beam to be
measured unambiguously. Further work is desired to experimentally
realize the scheme. The transformation method advanced in the
present work allows one to tailor a beam's wavefront by flat media
in a desirable way and should have practical implications for a
wide range of flat optical components. For simplicity we have only
considered the longitudinal transformation on wave. If taking into
account the transversal expansion/suppression to control the beam
dimension \cite{Emiroglu2010} or the azimuthal twist to control
polarization \cite{Kwon2008} as well, a more or full control on
wave will be achieved.

\begin{acknowledgements}
We gratefully acknowledge the referee for his insightful comments
and valuable suggestions. This work was supported in part by the
National Natural Science Foundation of China (No. 10847121,
10804029, 10904036, 61025024) and the Growth Program for Young
Teachers of Hunan University.
\end{acknowledgements}

\end{document}